\RequirePackage{fix-cm}
\documentclass[pdftex,twocolumn,epjc3]{svjour3}  
\RequirePackage[T1]{fontenc}
\RequirePackage{graphicx}
 \RequirePackage{mathptmx}      
 \RequirePackage{flushend}
\RequirePackage{latexsym}
\RequirePackage{multirow}
\RequirePackage[numbers,sort&compress]{natbib}

\RequirePackage{siunitx}

\RequirePackage[colorlinks,citecolor=blue,urlcolor=blue,linkcolor=blue]{hyperref}

\RequirePackage{lineno}
\RequirePackage{amsmath}
\RequirePackage{etoolbox} 
\RequirePackage{microtype}
\usepackage{siunitx}

\newcommand*\linenomathpatch[1]{%
  \cspreto{#1}{\linenomath}%
  \cspreto{#1*}{\linenomath}%
  \csappto{end#1}{\endlinenomath}%
  \csappto{end#1*}{\endlinenomath}%
}

\linenomathpatch{equation}
\linenomathpatch{gather}
\linenomathpatch{multline}
\linenomathpatch{align}
\linenomathpatch{alignat}
\linenomathpatch{flalign}

\journalname{Eur. Phys. J. C}

\begin{document}


\title{Direct Measurement of the $^{39}$Ar Half-life from 3.4 Years of Data with the DEAP-3600 Detector}

\author{
        P.~Adhikari~\thanksref{Carleton}
        \and
        R.~Ajaj~\thanksref{Carleton, Mcdonaldinst}
        \and
        M.~Alp\'{i}zar-Venegas~\thanksref{UNAM}
        \and
        P.-A.~Amaudruz~\thanksref{Triumf}
        \and
        J.~Anstey~\thanksref{Carleton, Mcdonaldinst}
        \and
        D.J.~Auty~\thanksref{Alberta}
        \and
        M.~Batygov~\thanksref{Laurentian}
        \and
        B.~Beltran~\thanksref{Alberta}
        \and
        M.A.~Bigentini~\thanksref{Carleton, Mcdonaldinst}
        \and
        C.E.~Bina~\thanksref{Alberta, Mcdonaldinst}
% ...
        (DEAP Collaboration)\thanksref{email}
}

\thankstext{deceased}{Deceased}
\thankstext{JMonroe}{Currently at the University of Oxford, Oxford,
OX1 3PU, United Kingdom}
\thankstext{TRPollman}{Currently at Nikhef and the University of Amsterdam, Science Park, 1098XG Amsterdam, Netherlands}
\thankstext{email}{deap-papers@snolab.ca}

\institute{
    Department  of  Physics,  University  of  Alberta,  Edmonton,  Alberta,  T6G  2R3,  Canada \label{Alberta}
% ...
\and
    Department  of  Physics,  Carleton  University,  Ottawa,  Ontario,  K1S  5B6, Canada \label{Carleton}
% ...
\and
    School of Natural Sciences, Laurentian University, Sudbury, Ontario, P3E 2C6, Canada \label{Laurentian}
% ...
\and
    Instituto de F\'{i}sica, Universidad Nacional Aut\'{o}noma de M\'{e}xico, A. P. 20-364, Ciudad de M\'{e}xico 01000, Mexico \label{UNAM}
% ...
\and
    TRIUMF, Vancouver, British Columbia, V6T 2A3, Canada \label{Triumf}
% ...
\and
    Arthur B. McDonald Canadian Astroparticle Physics Research Institute, Queen's University, Kingston, ON, K7L 3N6, Canada \label{Mcdonaldinst}
% ...
}

\date{Received: date / Accepted: date}

\maketitle
\begin{abstract}
The half-life of $^{39}$Ar is measured using the DEAP-3600 detector located 2~km underground at SNOLAB. Between 2016 and 2020, DEAP-3600 used a target mass of (3269 $\pm$ 24)~kg of liquid argon distilled from the atmosphere in a direct-detection dark matter search. Such an argon mass also enables direct measurements of argon isotope properties. The decay of $^{39}$Ar in DEAP-3600 is the dominant source of triggers by two orders of magnitude, ensuring high statistics and making DEAP-3600 well-suited for measuring this isotope's half-life. Use of the pulse-shape discrimination technique in DEAP-3600 allows powerful discrimination between nuclear recoils and electron recoils, resulting in the selection of a clean sample of $^{39}$Ar decays. Observing over a period of 3.4~years, the $^{39}$Ar half-life is measured to be $(302 \pm 8_{\rm stat} \pm 6_{\rm sys})$ years. This new direct measurement suggests that the half-life of $^{39}$Ar is significantly longer than the accepted value, with potential implications for measurements using this isotope's half-life as input.
\end{abstract}


\section{Introduction}
\label{introduction}
Argon is the third-most abundant gas in Earth's atmosphere, comprising about 0.93\% of the atmosphere by volume. The majority of this argon exists in the form of the stable isotopes $^{36}$Ar, $^{38}$Ar, and $^{40}$Ar with abundances of 0.0033, 0.0006, and 0.9960, respectively~\cite{Bohlke2014}. Interactions between cosmic ray-induced neutrons and this argon result in three long-lived radioisotopes, $^{39}$Ar, $^{37}$Ar, and $^{42}$Ar, with the dominant long-lived isotope being $^{39}$Ar~\cite{Benetti_2007,Calvo_2018}. 
The abundance of $^{39}$Ar was previously inferred to be \num{8.2e-16}~\cite{Ar40Ar39production, LOOSLI1968191, Benetti_2007} using the measured specific activity and half-life. This isotope $\beta$-decays to $^{39}$K with a Q-value of (565~$\pm$~5) keV~\cite{Ar39betaspectrum}. 
The most recent measurement for the specific activity of $^{39}$Ar was performed by the DEAP-3600 collaboration with a value of (0.964~$\pm~0.001_{\rm stat}~\pm~0.024_{\rm sys}$) Bq/kg$_{\rm atmAr}$~\cite{specificactivityAr39}.
Understanding the half-life of $^{39}$Ar is important in areas such as radioisotope dating, radiochemistry, radiometrology, geochronology, and nuclear physics measurements~\cite{geo1, geo2, radio1}. 

Several measurements of the half-life of $^{39}$Ar exist, with the first estimate of 4 minutes made in 1937~\cite{39Ar1937,39Ar1948}. This estimate was contradicted by Brosi et al. in 1950 who, while investigating the $\beta$-spectrum for this isotope, estimated a lower limit of the half-life of greater than 15 years~\cite{Ar39betaspectrum}.
In 1952 they presented a half-life and mass measurement of $^{39}$Ar in Zeldes et al.~\cite{39Ar1952} using argon produced from four samples of potassium chloride which had been bombarded in a nuclear reactor for about one year: the half-life was calculated for three of the samples independently by comparing isotopic ratios from mass spectrometry measurements to the fourth sample whose decay rate was measured in a magnetic lens beta-ray spectrometer. Their results ranged from 240 to 290 years and an average value was reported as \SI[separate-uncertainty]{265(30)}{years}.

The most widely referenced value for the half-life of $^{39}$Ar is (269 $\pm$ 3$_{\rm stat}$ $\pm$ 8$_{\rm sys}$) years; this value was estimated by Stoenner et al. in 1965~\cite{39Ar1965} from the isotopic ratio of $^{39}$Ar to $^{37}$Ar and the measured disintegration rates in 10 samples. Their reported half-life measurement is the average from the 10 samples with the individual measured values ranging from 253 to 288 years. An earlier publication by Stoenner et al. in 1960~\cite{Stoenner1960} gave an $^{39}$Ar half-life of 325~years but did not provide an uncertainty. For the purpose of comparison, uncertainties on their measured rates of $\sim$5\% were propagated to their half-life measurement giving an estimated half-life with statistical uncertainty of \SI[separate-uncertainty]{325(16)}{years}.

A method to measure the half-life of $^{39}$Ar through mass spectrometry by recording the $^{39}$Ar to $^{38}$Ar isotopic ratios in a double-spike over a period of 31.8 years was published by Baksi et al. in 1996~\cite{39Ar1996}. Their measurement of the $^{39}$Ar half-life is reported as \SI[separate-uncertainty]{276(3)}{years}. 
While they recognized the ``use of measurements made at different laboratories using different instrumentation'' without taking this into account as a source of uncertainty, and given the spread of their three data points, it is possible that the systematic uncertainty they reported may be an underestimate. The Nuclear Data Sheets (NDS) from 2018, produced by the National Nuclear Data Center at Brookhaven National Laboratory, re-evaluated the Stoenner et al. 1965 result using an updated $^{37}$Ar half-life and give the $^{39}$Ar half-life as \SI[separate-uncertainty]{268(8)}{years} \cite{39Ar2018}.

To date there have been no direct measurements of the $^{39}$Ar half-life by continuously observing its decay curve. The DEAP-3600 detector can directly observe the decay of $^{39}$Ar, which is the dominant source of triggers in the detector. This paper presents a measurement of the $^{39}$Ar half-life by observing changes in the event rate due to the decay of this isotope. Existing measurements are compiled along with this result and presented in Table~\ref{tab:lifetimeSummary}.

\begin{table}
\centering
\label{tab:lifetimeSummary}
\caption{A comparison of existing $^{39}$Ar half-life measurements with this result.}
\resizebox{\columnwidth}{!}{
\begin{tabular}{lll}
\hline\noalign{\smallskip}
\textbf{Measurement} & \textbf{Half-life} & \textbf{Method}\\
& \textbf{[years]} &\\
\hline\noalign{\smallskip}
Zeldes et al. (1952) & 265 $\pm$ 30 & Isotopic ratios of\\
& & argon samples using\\
& & mass spectrometry\\
Stoenner et al. (1960) & 325 $\pm$ 16$^{\dagger}_{\rm stat}$ & Activity ratios of\\
& & $^{39}$Ar to $^{37}$Ar\\
Stoenner et al. (1965) & 269 $\pm$ 3$_{\rm stat}$ $\pm$ 8$_{\rm sys}$ & Activity ratios of\\
& & $^{39}$Ar to $^{37}$Ar\\
Baksi et al. (1996) & 276 $\pm$ 3$_{\rm sys}$ & Isotopic ratios of\\
& & $^{39}$Ar to $^{38}$Ar in a\\
& & double-spike using\\
& & mass spectrometry\\
NDS (2018) & 268 $\pm$ 8$_{\rm sys}$ & Re-evaluation of the\\
& & Stoenner et al. (1965)\\
& & result with an updated\\
& & $^{37}$Ar half-life\\
DEAP-3600 (2025) & 302 $\pm$ 8$_{\rm stat}$ $\pm$ 6$_{\rm sys}$ & Direct observation\\
& & of the decay curve\\
\hline\noalign{\smallskip}
\end{tabular}
}
$^{\dagger}$Estimated from uncertainty on the measured count rates.
\end{table}

\begin{figure} 
    \includegraphics[width=\columnwidth]{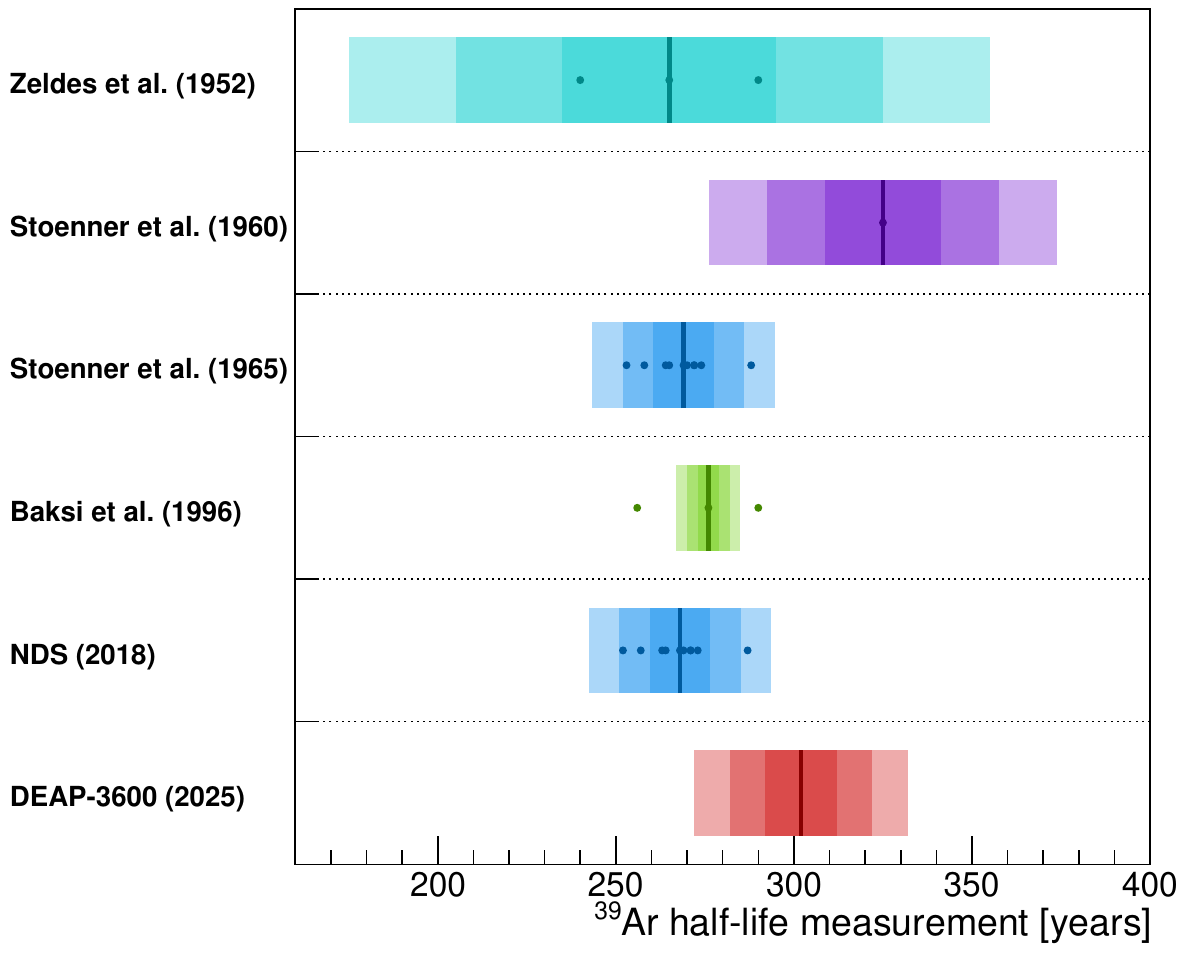}
    \centering
    \caption{A comparison of existing $^{39}$Ar half-life measurements with this work. The points shown for Zeldes et al. (1952)~\cite{39Ar1952}, Stoenner et al. (1960)~\cite{Stoenner1960}, Stoenner et al. (1965)~\cite{39Ar1965}, Baksi et al. (1996)~\cite{39Ar1996}, and NDS (2018)~\cite{39Ar2018} denote the individual measurements which were reported. The vertical line represents either the average value or the single reported value. The shaded boxes indicate the 1$\sigma$, 2$\sigma$, and 3$\sigma$ error bands with Stoenner et al. (1960) showing the statistical error, Baksi et al. (1996) and NDS (2018) showing the systematic error, and the others showing the combination of their statistical and systematic uncertainties.}
    \label{fig:lifetimeComp}
\end{figure}

The DEAP-3600 detector is described in Section~\ref{sec:detector}. Section~\ref{sec:dataset} provides a description of the dataset along with the event selection procedure and detector livetime calculation.  The model used to fit the data and measure the $^{39}$Ar half-life is derived in Section~\ref{sec:fitModel}, and the parameters of this model are described in Section~\ref{sec:FitModelValues}.  Systematic uncertainties are described in Section~\ref{sec:systematic_uncertainties}, and the result is given with concluding statements in Section~\ref{sec:conclusion}. 

\section{The DEAP-3600 Detector}
\label{sec:detector}

The DEAP-3600 detector is located 2 km underground at SNOLAB in Sudbury, Canada, and was designed to observe scintillation light produced when energy is deposited in the target material of (3269 $\pm$ 24)~kg of ultra-pure liquid argon (LAr)~\cite{specificactivityAr39}. A detailed description of the detector can be found in Ref.~\cite{Design_DEAP}.

\begin{figure}
\centering
\includegraphics[width=\columnwidth]{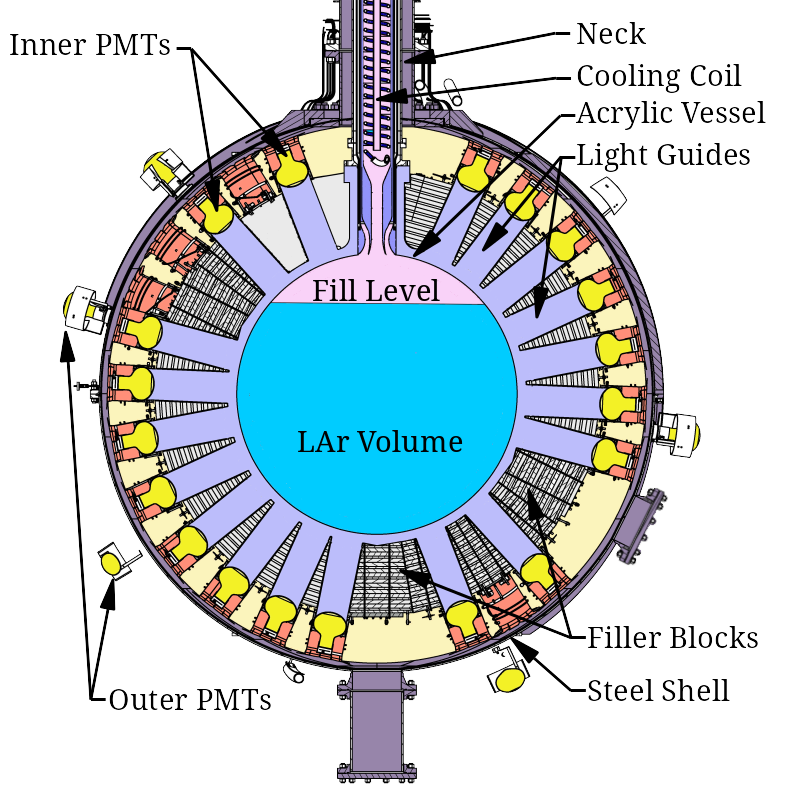}
  \caption{A cross-section of the DEAP-3600 detector components which are located within the cylindrical water tank.}
  \label{fig:detector}
\end{figure}

The DEAP-3600 detector consists of a spherical acrylic vessel (AV) with a radius of 85 cm which contains the LAr. Surrounding the AV are 255 inward-facing Hammamatsu R5912 Photomultiplier Tubes (PMTs) which observe and record scintillation light from the LAr. During data taking the AV was partially filled with the LAr level approximately 55~cm above the equator. This LAr remained hermetically sealed within the AV for the duration of data taking. A 3~$\mu$m thick layer of tetraphenyl butadiene (TPB) coats the inner surface of AV and is present to shift the ultraviolet scintillation light at 128~nm into the visible light spectrum at 420~nm~\cite{tpb_2013}. Attached to the top of the AV is a cylindrical acrylic neck. This neck surrounds a liquid nitrogen filled cooling coil which condenses gaseous argon (GAr) in the upper portion of the AV. The AV is situated within a spherical stainless steel shell which is continuously flushed with radon-scrubbed nitrogen gas. The steel shell is submerged in a cylindrical tank containing ultra-pure water. This tank, along with 48 additional PMTs on the outer surface of the steel shell, acts as a muon veto system by detecting Cherenkov radiation produced by muons passing through the water. A schematic of the detector is shown in Figure~\ref{fig:detector}.
 
Within the data acquisition system (DAQ) each PMT is connected to a single channel on a custom-built signal conditioning board (SCB) which decouples each PMT's high voltage and shapes the signals. Outputs from the SCBs are transmitted to high-gain (CAEN V1720) and low-gain (CAEN V1740) digitizers which convert continuous analogue signals into discrete digital signals using analogue-to-digital converters. The summed output from each SCB is also transmitted to a digitizer and trigger module (DTM) which makes triggering decisions based on two rolling charge integrals: a narrow integral $Q_n$ over a 177~ns window and a wide integral $Q_w$ over a 3.1~$\mu$s window. The promptness of the signal is computed from the $Q_n/Q_w$ ratio. Five trigger regions are defined based on these three variables and a prescaling factor of 100 is applied to events in the energy range of $Q_w\approx$~{(50 to 565)~$\text{keV}_\text{ee}$}, in the low $Q_n/Q_w$ region. This prescaling mainly affects $^{39}$Ar decays and reduces the statistics by storing PMT waveforms for only 1 out of every 100 triggers. Roughly \num{2.7e6} $^{39}$Ar decays remain from a 24-hour period after this prescaling. The timestamp for every trigger is recorded, including those which are prescaled.

Each digitizer channel records PMT waveforms for 16~$\mu$s, including a pre-trigger window of 2.4~$\mu$s. Each PMT's observed charge is integrated over a window of [-28,~10000]~ns relative to the event time. The PMTs are calibrated on a daily basis~\cite{spe} and a PMT charge response model is used to calculate the number of photoelectrons (PEs) detected in each identified pulse which provides the energy estimator for data. The DAQ is operated using \texttt{MIDAS}~\cite{MIDAS_DAQ}, and the data are analyzed with the \texttt{RAT} software framework~\cite{RAT_framework} which is based on \textsc{Geant4}~\cite{AGOSTINELLI2003250} and \texttt{ROOT}~\cite{BRUN199781}. 

The pulse-shape discrimination (PSD) parameter $F_{\rm prompt}$ \cite{DEAP_PSD} is defined for each event as the ratio of prompt to total charge, 
\begin{equation}
{F_{\rm prompt}} = \frac{\sum_{t=-28~{\rm ns}}^{t={150~{\rm ns}}} Q(t)}{\sum_{t=-28~{\rm ns}}^{t={10~{\rm \mu s}}} Q(t)}
\label{eq:Fprompt}
\end{equation}
where $Q(t)$ is the pulse charge in PE and $t$ is the pulse time relative to the trigger time. $F_{\rm prompt}$ is sensitive to the ratio of singlet to triplet excimer states in LAr, with lifetimes of 6~ns~\cite{Heindl_2010} and approximately 1.5~$\mu$s~\cite{pulseshapes}, respectively. This ratio is significantly different for electron recoils and nuclear recoils. This technique provides powerful discrimination between events in the nuclear recoil band (NRB) centered at \mbox{$F_{\rm prompt}\approx0.70$} and the electron recoil band (ERB) centered at \mbox{$F_{\rm prompt}\approx0.25$}. Figure~\ref{fig:fprompt_vs_PE_ROI} shows $F_{\rm prompt}$ over a wide PE range.

Another parameter of interest for this analysis, $F_{\rm maxPE}$, gives the fraction of the total PEs seen by the PMT which observed the most PEs in an event. A high value for this parameter indicates that the majority of the signal from an event was observed by a single PMT. This would be the case when Cherenkov radiation is produced in a light guide, for example.

\section{Dataset}
\label{sec:dataset}

The dataset for this analysis is divided into discrete runs which were taken over a period of 3.4 years. Individual runtimes $T_{\rm run}$ vary in the range of [$\mathcal{O}(\SI{1}{minute})$, \SI{\sim2}{days}] but a typical run is roughly 22 hours long. A requirement that $T_{\textrm{run}}\geq$ 4.85 hours is made to ensure sufficient statistics for the light yield corrections described in Section \ref{sec:lycorr}. Runs selected for analysis meet several criteria including the stability of the LAr cooling system, the charge distribution of the PMTs, and the efficiency of the trigger.

\begin{figure} 
    \includegraphics[width=\columnwidth]{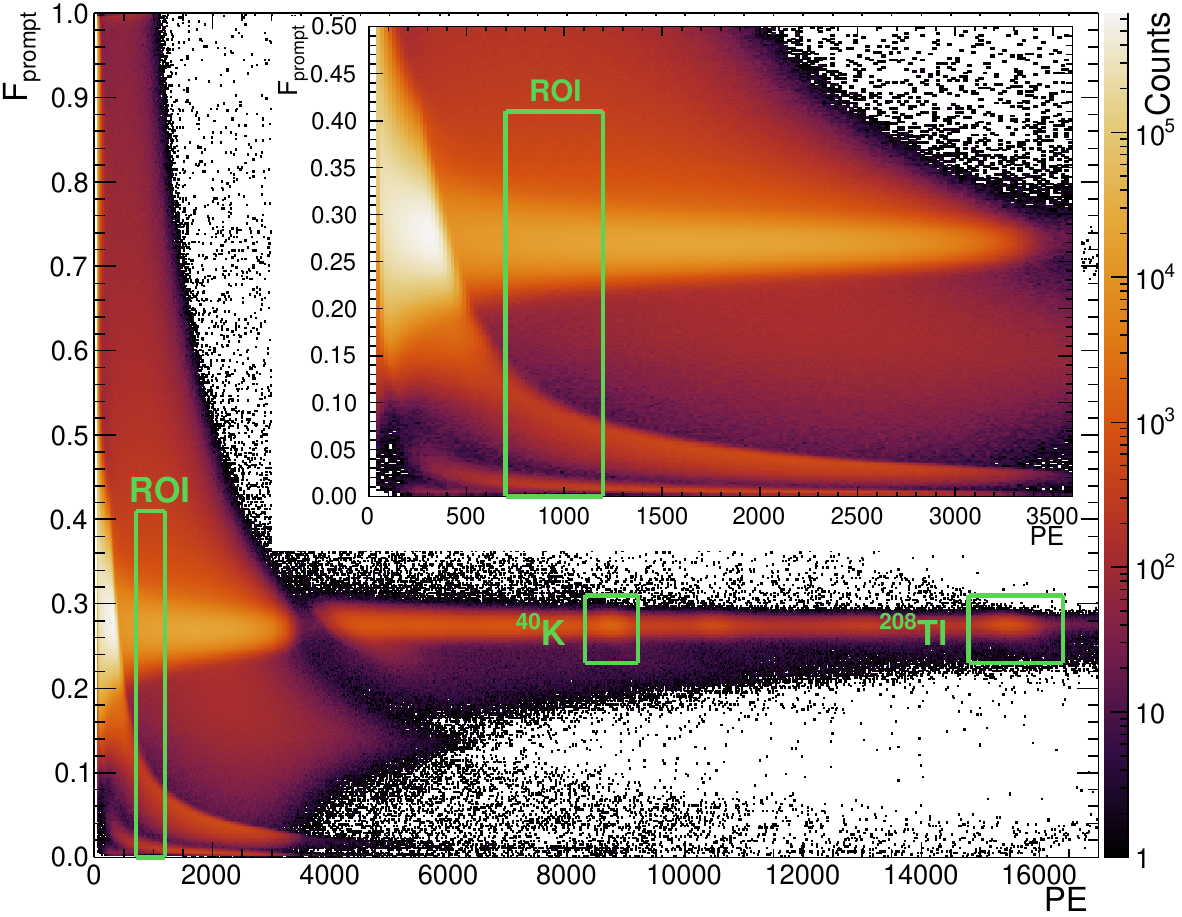}
    \centering
    \caption{Histogram of $F_{\rm prompt}$ versus the number of observed PE. The ROI is highlighted in the left-most box for $\sim$20 days of data. All the events in the ROI are within the prescaled region.  The regions labeled $^{40}$K and $^{208}$Tl highlight the $\gamma$-peak positions of those isotopes. The events within the ROI at $F_{\rm prompt}$ < 0.15 are due to pile-up and are accounted for in the analysis.}
    \label{fig:fprompt_vs_PE_ROI}
\end{figure}

\begin{figure} 
    \includegraphics[width=\columnwidth]{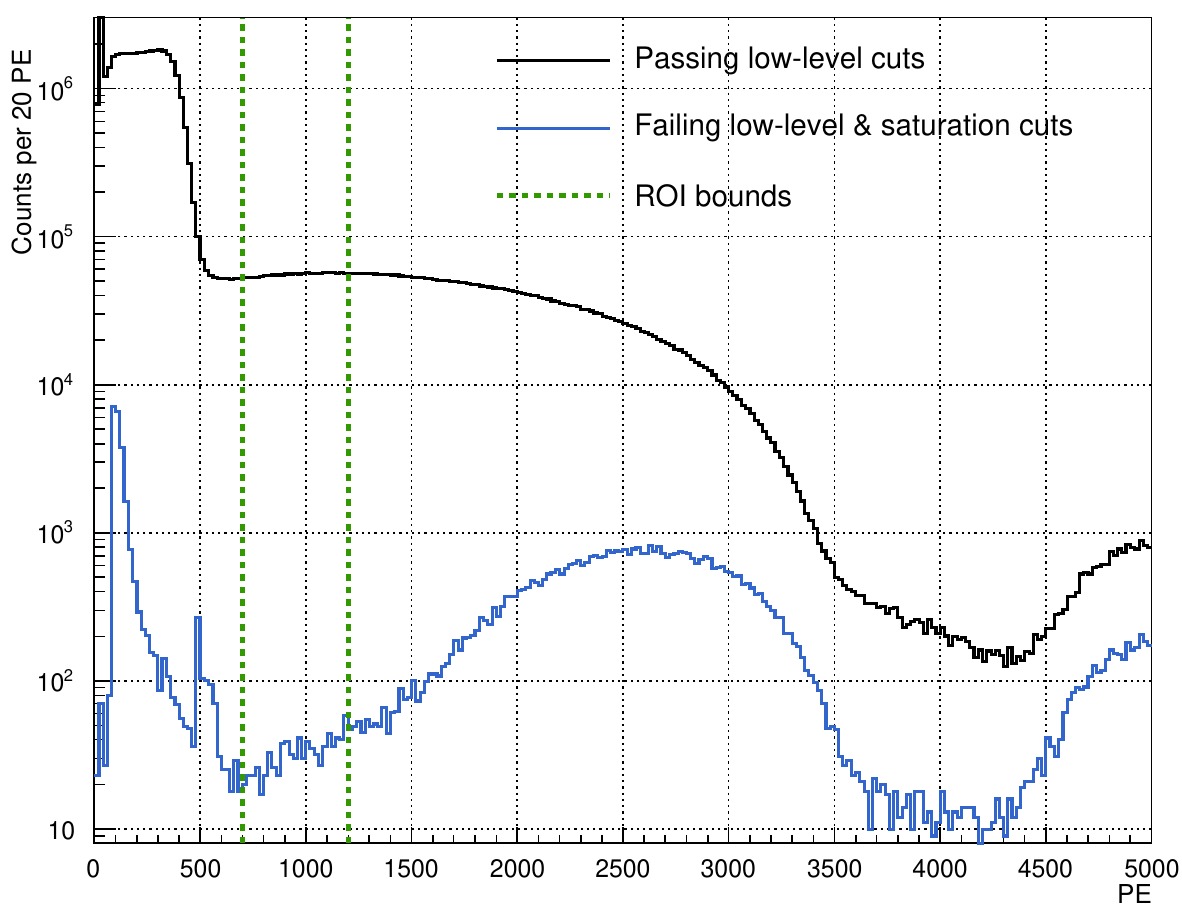}
    \centering
    \caption{The PE spectrum of events with $F_{\rm prompt}$ $\leq$~0.41, showing separately the subset of events failing low-level cuts or where PMT saturation effects are present. The prescaling region, seen as a sharp drop at around 500 PE and a rise at around 4500 PE, and the events where PMT saturation begins to increase at around 1400 PE can be seen. These two boundaries inform the ROI selection with range 700--1200~PE.}
    \label{fig:PE_vs_different_calcut}
\end{figure}

A subset of the full energy spectrum between 700 and 1200~PE (approximately \SIrange[]{115}{195}{keV}) is selected for analysis. The region of interest (ROI) is highlighted in Figure~\ref{fig:fprompt_vs_PE_ROI}. The lower limit at 700 PE is selected to be well above the prescaling boundary at around 500 PE. The upper limit at 1200 PE is selected to reject any systematic effects resulting from saturation of the PMTs. These effects can be seen in Figure \ref{fig:PE_vs_different_calcut}. The limit of $F_{\rm prompt}$ $\leq$ 0.41 is made to reject events outside of the ERB.

Within the ROI the PE spectrum is comprised of single $^{39}$Ar events, double and triple $^{39}$Ar pile-up events (uncorrelated coincidences in which multiple signals from $^{39}$Ar decays occur within the same trigger window), and events which contain the pile-up of $^{39}$Ar with Cherenkov radiation. A final contribution comes from the low-energy tails of $\gamma$-rays resulting from the decay of daughter products of long-lived radioisotopes in the detector components. This $\gamma$-ray component, however, is relatively featureless in the ROI and is considered to be a constant background.

\begin{figure} 
    \includegraphics[width=\columnwidth]{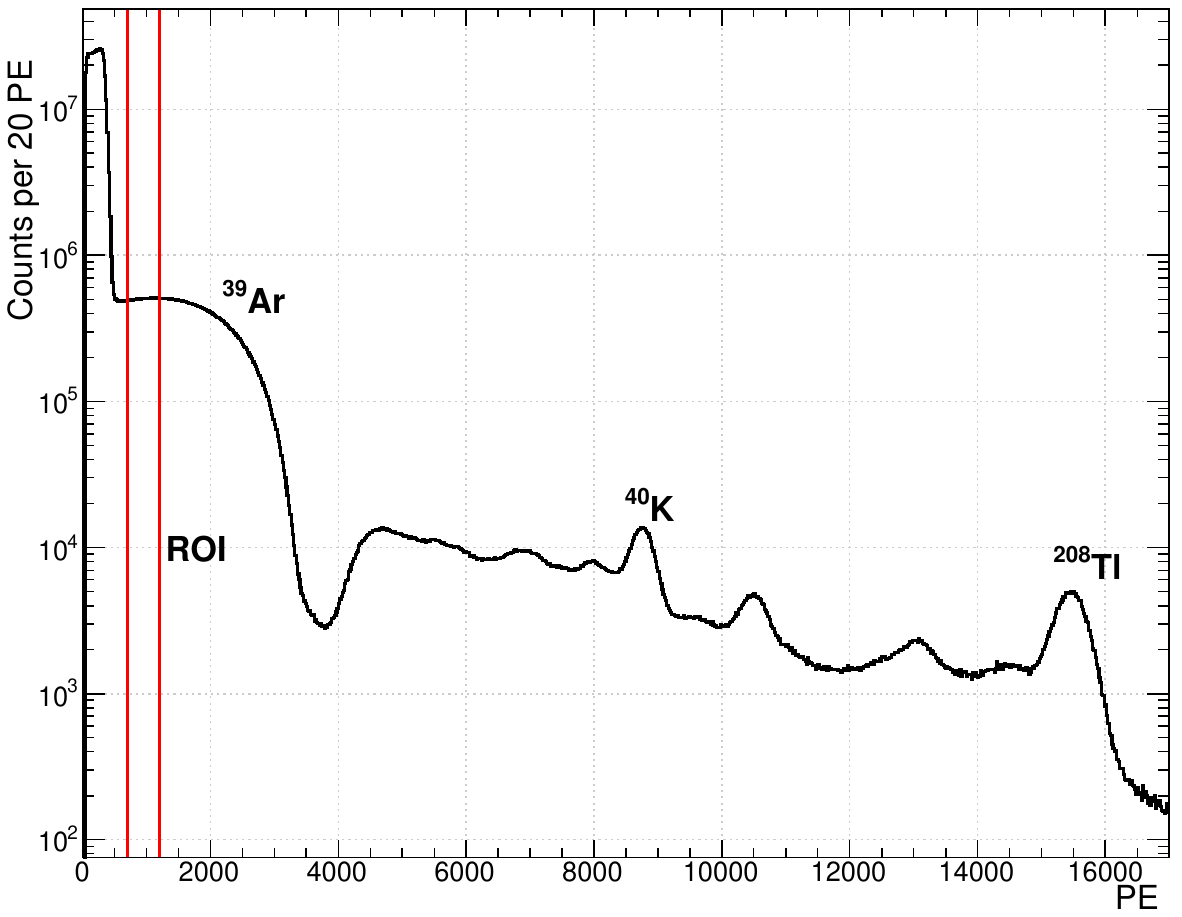}
    \centering
    \caption{The PE spectrum of events with $F_{\rm prompt}$ $\leq$~0.41. The three light yield calibration sources are highlighted. The lower and upper prescaling bounds, at $\sim$500 PE and $\sim$4000 PE respectively, are clearly seen.  This spectrum represents $\sim$20 days worth of data.}
    \label{fig:1Dspectrum}
\end{figure}

A data-cleaning cut is applied to data which rejects events occurring within $\delta t_{\rm cut}$ = 70~$\mu$s of an earlier event, where late light from the earlier event may leak into the trigger window. 
The number of events removed with this cut is $N_{\rm DCcut}$, with $\delta t_k\leq$70~$\mu$s removed for each event $k$. Further low-level cuts are made to reject any trigger sources other than LAr scintillation such as internal calibration and muon veto triggers, removing $\delta t_{\rm cut}$ for each of the total $N_{\rm LLcut}$ events.  The number of remaining physics triggers $N_{\rm phys}$ each introduce additional deadtime of $\delta t_{cut}$. The livetime $T_{\rm live}$ for a given run is calculated as follows:
\begin{equation}
T_{\rm live} = T_{\rm run} - \sum_{k = 1}^{N_{\rm DCcut}} \delta t_{k} - N_{\rm LLcut} \cdot \delta t_{\rm cut} - N_{\rm phys} \cdot \delta t_{\rm cut}.
\label{eq:tlive}
\end{equation}

A final requirement for the remaining $N_{\rm phys}$ events is that $F_{\rm maxPE}$ is less than 0.4. A parameter in data which is a count of the sub-events within a trigger window is used to identify pile-up events; no cut is applied on this parameter for the nominal fit to data, however a cut is applied when evaluating the double and triple $^{39}$Ar pile-up event selection efficiencies described in Section~\ref{sec:evEffic}.

\subsection{Correction for Light Yield Variations}
\label{sec:lycorr}
\begin{figure} 
\includegraphics[width=\columnwidth]{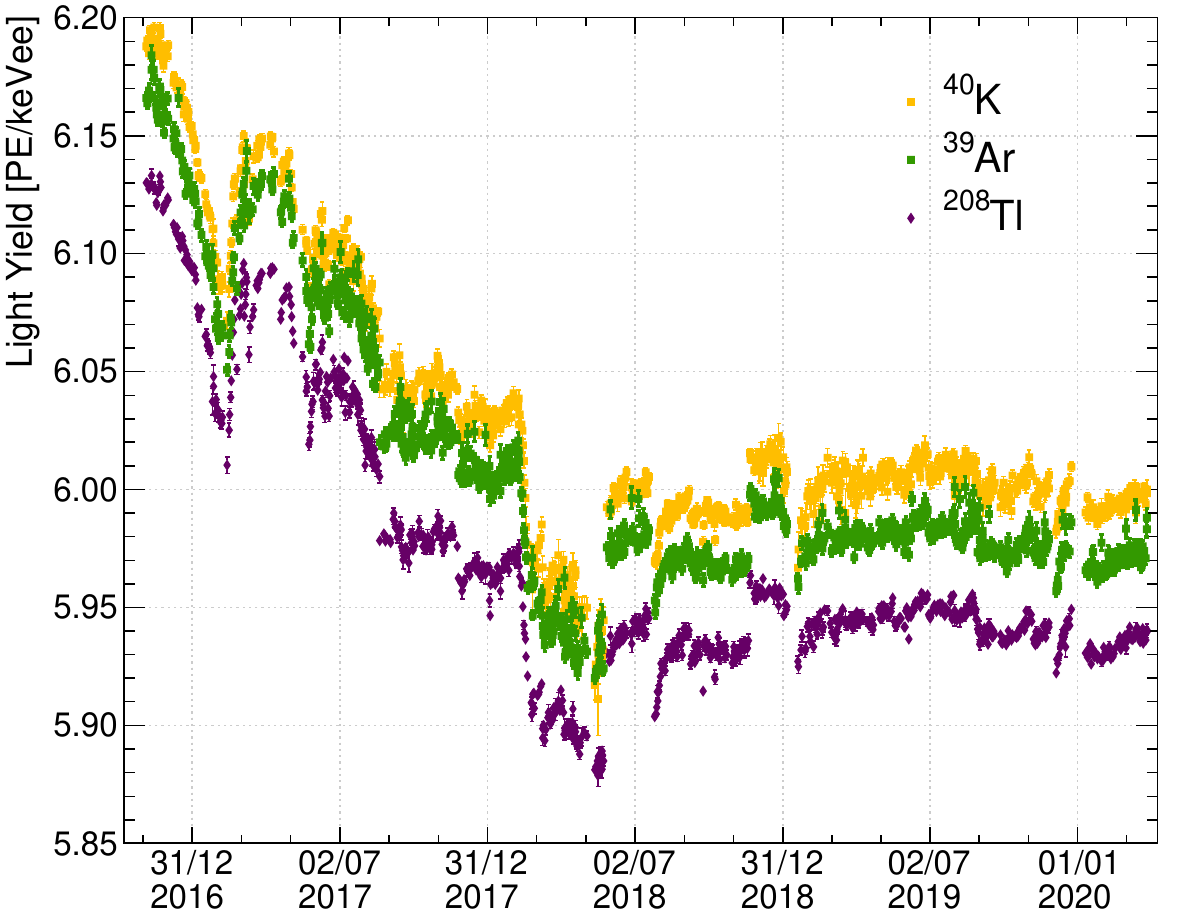}
\caption{The light yield as measured from three calibration sources for all runs in the dataset. Variations over time can be seen, though the general trends are similar for each source. The error bars shown are the statistical uncertainties from the fits to each source.}
\label{fig:LYvar}
\end{figure}

The light yield, a ratio of the number of PEs which are observed by the PMTs to the amount of energy deposited in the LAr, is measured in the ERB from three calibration sources. The $\gamma$-peaks of $^{40}$K (at 1460~keV, or $\sim$8,750~PE) and $^{208}$Tl (at 2615~keV, or $\sim$15,500~PE) are fit with Gaussian distributions and the mean PE value of each peak is divided by their respective peak energies. The third source is measured by fitting the $^{39}$Ar spectrum with the convolution of the $\beta$-spectrum model described in Ref.~\cite{kostensalo} and a detector response model which includes a light yield parameter. Each of these sources are shown for data in Figure~\ref{fig:1Dspectrum}. Figure \ref{fig:LYvar} shows the run-by-run light yield measurements where the variations over the length of the dataset can be seen.

These variations in the light yield affect the fraction of events within the ROI across the dataset by stretching or compressing the PE spectrum. To correct for these variations, the fractional change in the average light yield ratio is calculated for each run $j$ as the average light yield ratio $Y_j$ for that run to that of the first run $Y_0$. Both $Y_j$ and $Y_0$ are the arithmetic means of the three different calibration sources for their respective runs. The PE value of each event in a run is then corrected by a factor of $Y_j/Y_0$.

\section{Model of $^{39}$Ar Trigger Rates}
\label{sec:fitModel}

The model used to fit the $^{39}$Ar trigger rate includes components which account for single $^{39}$Ar events and pile-up events. A trigger may consist of a single $^{39}$Ar event, multiple $^{39}$Ar pile-up events, or $^{39}$Ar-Cherenkov pile-up events. The trigger rate in the ROI is given by 
\begin{equation} 
    \label{eq:rate2}
    R(t) = \sum_{i=1}^{4}R_i(t)\eta_{i}f_{i}\epsilon_{i},
\end{equation}
where $R(t)$ is the total trigger rate at time $t$, $R_i$ are the rates for the various event types with \mbox{$i$ = 1} for single $^{39}$Ar, \mbox{$i$ = 2} and 3 for double and triple $^{39}$Ar pile-up respectively, and \mbox{$i$ = 4} for $^{39}$Ar-Cherenkov pile-up. The $f_i$ terms are the fraction of each spectrum that falls within the ROI, the $\epsilon_{i}$ terms are the event selection efficiencies. The $\eta_{i} = 1/N$ terms account for the number of uncorrelated events $N$ observed in each trigger with \mbox{$\eta_{1}$ = 1} for single $^{39}$Ar decays, \mbox{$\eta_{2}$ = 1/2} for double $^{39}$Ar pile-up, \mbox{$\eta_{3}$ = 1/3} for triple $^{39}$Ar pile-up, and \mbox{$\eta_{4}$ = 1/2} for $^{39}$Ar-Cherenkov pile-up.

Including a constant ERB $\gamma$ background trigger rate $R_{\rm bg}$ (see Section~\ref{ER_bg_deap}), Eq.~\ref{eq:rate2} becomes
\begin{multline}    
    \label{eq:rate3}
    R(t) = R_1(t)f_{1}\epsilon_{1} ~+~ \frac{1}{2}R_{2}(t)f_{2}\epsilon_{2} ~+\\
    \frac{1}{3}R_{3}(t)f_{3}\epsilon_{3} ~+~ \frac{1}{2}R_{4}(t)f_{4}\epsilon_{4} ~+~ R_{\rm bg}.
\end{multline}

The rate of single $^{39}$Ar decays in a time interval \mbox{$\delta t$ = 10~$\mu$s} is given by the product of the decay rate with the probability $P(0)$ that there is no other decay in that interval as determined from Poisson statistics.  This probability is given by
\begin{equation}
\label{eq:poissonAr}
P(0) = e^{-R_{\rm 39Ar}(t)\delta t},
\end{equation}
where $R_{\rm 39Ar}(t)$ is the $^{39}$Ar trigger rate. The trigger rate for single $^{39}$Ar decays is then
\begin{equation}  
\label{eq:poissonArSingle}
R_1(t) = R_{\rm 39Ar}(t)P(0) = R_{\rm 39Ar}(t)e^{-R_{\rm 39Ar}(t)\delta t}.
\end{equation}

The rates for double and triple $^{39}$Ar pile-up triggers can similarly be determined from Poisson statistics, where the probability $P(1)$ for one other decay in the interval given by
\begin{equation}
    P(1) = R_{\rm 39Ar}(t)e^{-R_{\rm 39Ar}(t)\delta t}\delta t
\end{equation}
and the probability $P(2)$ for two other decays in the interval given by
\begin{equation}
    P(2) = \frac{1}{2}R^{2}_{\rm 39Ar}(t)e^{-R_{\rm 39Ar}(t)\delta t}\delta t^{2}.
\end{equation}
The trigger rates are then given by
\begin{equation}
    R_2(t) = R_{\rm 39Ar}(t)P(1) = R^{2}_{\rm 39Ar}(t)e^{-R_{\rm 39Ar}(t)\delta t}\delta t,\\
\end{equation}
and
\begin{equation}
    R_3(t) = R_{\rm 39Ar}(t)P(2) = \frac{1}{2}R^{3}_{\rm 39Ar}(t) e^{-R_{\rm 39Ar}(t)\delta t}\delta t^{2}.
\end{equation}
Pile-up of $^{39}$Ar with Cherenkov radiation is given by
\begin{equation}
\label{eq:ArCherPileupRate}
R_{4}(t) = R_{1}(t)R_{\rm Chv}(t)e^{-R_{\rm Chv}(t)\delta t}\delta t,
\end{equation}
where $R_{\rm Chv}(t) = R_{\rm Chv}$ is the rate of Cherenkov triggers, observed to be constant across the dataset. 

The $^{39}$Ar trigger rate can be expanded to
\begin{equation} 
\label{eq:decay}
R_{\rm 39Ar}(t) = R_{\rm 39Ar}\ e^{-t/\tau_{\rm 39Ar}},
\end{equation}
where $R_{\rm 39Ar}$ is taken at the start of the dataset, and $\tau_{\rm 39Ar}$ is the mean lifetime of $^{39}$Ar. These are the two parameters that are free-floating in the fits.

The total trigger rate in Eq.~\ref{eq:rate3} then becomes
\begin{align} 
\label{eq:model}
\begin{split}
R(t) &= e^{-R_{\rm 39Ar}(t)\delta t}\biggl[R_{\rm 39Ar}(t) f_{1}\epsilon_{1}\\
&+~\frac{1}{2}R_{\rm 39Ar}^2(t)\delta t f_{2}\epsilon_{2}+\frac{1}{6}R_{\rm 39Ar}^3(t)\delta t^2 f_{3}\epsilon_{3}\\
&+~\frac{1}{2}R_{\rm 39Ar}(t)R_{\rm Chv} e^{-R_{\rm Chv}\delta t}\delta t f_{4}\epsilon_{4}\bigg]~+~R_{\rm bg}.
\end{split}
\end{align}
The mean lifetime measured with this model can then be converted into a half-life: $T_{1/2}~=~\ln(2)\tau_{\rm 39Ar}$. The parameters in Eq.~\ref{eq:model} are further described in Section~\ref{sec:FitModelValues} and summarized in Table~\ref{tab:Summary_table}.
\section{Fit Model Parameters}
\label{sec:FitModelValues}
This section provides details of the parameters of the fit model described in Sec.~\ref{sec:fitModel}, including how each parameter value is determined. Two sets of Monte Carlo (MC) simulations were performed to estimate these parameters. The first set of toy MC takes the analytical $\beta$-spectrum of $^{39}$Ar given in Ref.~\cite{kostensalo} and convolves it with a detector response model to estimate detector effects. This model uses a Gaussian response with mean $\mu$ given by
\begin{align} 
\mu = <N_{DN}> +~Y\cdot E,
\label{eq:energy_response}
\end{align}
where <$N_{DN}$> is the average number of PE produced by dark noise and stray photons in the trigger window, $Y$ is the light yield of the detector, and $E$ is the energy of an event. The average value of the dark noise has been measured to be (1.1 $\pm$ 0.2) PE~\cite{DEAP_231}. A toy MC simulation uses this model to estimate the fraction of different pile-up $^{39}$Ar events, $f_{2}$ and $f_{3}$, which fall within the ROI by randomly selecting events and adding their PE values to generate pile-up spectra. These pile-up spectra are also used to estimate the values of $\epsilon_{2}$ and $\epsilon_{3}$. The pile-up spectrum of $^{39}$Ar with low-energy Cherenkov radiation (described in Section~\ref{sec:ChvRad}) is generated using a similar toy MC to estimate the values of $f_{4}$ and $\epsilon_{4}$.

The second set of MC simulations used \texttt{RAT} to estimate detector effects. Simulations of $\gamma$-rays emitted by $^{40}$K and $^{208}$Tl in the PMT glass were performed at various trigger thresholds. The efficiency of the cuts included in the analysis is calculated from these simulated samples. Additionally, \texttt{RAT} was used to verify the fit model parameters evaluated from the toy MC simulations.

\subsection{Rate of Cherenkov Radiation}
\label{sec:ChvRad}
Cherenkov radiation is generated in acrylic and PMT glass by electrons with kinetic energies above thresholds of 175 keV and 126 keV, respectively. MC simulations were performed with \texttt{RAT} to obtain the full Cherenkov energy spectrum from $^{40}$K and $^{208}$Tl decays in the PMTs with the detector threshold set to values of \mbox{0, 1, and 3~PE}. The Cherenkov spectrum is also measured from data using a trigger threshold of 3~PE. The simulated spectra match data well. The Cherenkov spectrum from all other data taken at the nominal detector threshold is normalized to the spectrum obtained from the low-threshold run. Weekly averages for the Cherenkov rates are stable across the dataset, as expected, and small variations observed are included as a systematic uncertainty. The trigger rate due to Cherenkov radiation is determined to be $R_{\rm Chv} =$~\SI[separate-uncertainty]{538(4)}{Hz}.

\subsection{Event Selection Efficiencies}
\label{sec:evEffic}
The efficiency of the $F_{\rm prompt}$ and $F_{\rm maxPE}$ cuts used to select events was investigated across all runs in the dataset. The efficiencies are calculated as the ratio of triggers passing a cut to the total number of triggers before that cut within the PE range of the ROI. The product of efficiencies for all the cuts is determined on a run-by-run basis.

The efficiency for the selection of single $^{39}$Ar decay triggers is 100\% in both the MC samples and data. The efficiency for selecting double $^{39}$Ar pile-up triggers in MC is roughly 91\%. In data selected with the additional requirement that exactly two sub-events are detected, this efficiency is roughly 85\%; applying the same additional cut to MC results in the same value. The efficiency in data is corrected to account for the effect of this additional cut by adding the difference between the value determined in MC and the average pre-correction efficiency in data to the efficiency in data for each run. The corrected mean efficiency for selecting double $^{39}$Ar pile-up triggers in data is then \SI[separate-uncertainty]{0.9099(0.0033)}{}. For triple $^{39}$Ar pile-up trigger efficiencies, in a similar method but selecting events with exactly 3 sub-events, the difference between data and MC is taken into account and the corrected mean efficiency is \SI[separate-uncertainty]{0.860(0.039)}{}. The efficiency for selecting $^{39}$Ar-Cherenkov pile-up is evaluated using MC to be 100\%, consistent with the amount of Cherenkov light being small compared to scintillation from the single $^{39}$Ar decay in that event. 

\subsection{ERB $\gamma$ Background Rate} 
\label{ER_bg_deap}
The PE spectrum for electron recoils in DEAP-3600 is described in detail in Ref.~\cite{ER_bg_paper}. The trigger count for these backgrounds in the ROI is determined by subtracting the MC $^{39}$Ar spectrum from the total MC ERB sum. The difference is then normalized to the livetime of the dataset and the rate is determined to be $R_{\rm bg}$ = \SI[separate-uncertainty]{1.65(0.31)}{Hz}.

\begin{table*}
\caption{List of parameters used in the model of $^{39}$Ar trigger rates (Sections \ref{sec:fitModel}, \ref{sec:FitModelValues}) along with their uncertainties. The uncertainties due to the event selection efficiencies for each event type are given as a combined value. The uncertainties are estimated with respect to the measured mean lifetime and converted to half-life values. The character `--' indicates an uncertainty of less than 0.01 years which is considered negligibly small.}
\centering
\begin{tabular*}{\textwidth}{@{}lllll@{}}
\hline
\textbf{Parameter} & \textbf{Symbol} & \textbf{Value} & \textbf{Constraints} & \textbf{Absolute uncertainty}\\
& & & & \textbf{on $T_{1/2}$ [years]}\\
\hline\noalign{\smallskip}
$^{39}$Ar trigger rate at dataset start & $R_{^{39}\textrm{Ar}}$ &  & Free-floating & \\
$^{39}$Ar lifetime & $\tau_{^{39}\textrm{Ar}}$ & & Free-floating\\
Cherenkov trigger rate & $R_{\textrm{Chv}}$ & (538 $\pm$ 4)~Hz & Fixed & --\\ 
ERB background rate & $R_{\textrm{bg}}$ & (1.65 $\pm$ 0.31)~Hz & Fixed & 0.15\\ 
Livetime & $T_{\rm live}$ & 1.80 years & Fixed & --\\
Bin width for trigger rate averages & & 7 days & Fixed & 1.0\\
\hline
Fraction of single $^{39}$Ar spectrum in the ROI & $f_{1}$ & 0.21 $\pm$ 0.05 & Fixed & 0.21\\
Cut efficiency for single $^{39}\textrm{Ar}$ & $\epsilon_{1}$ & 1  & Fixed & --\\
\hline
Fraction of double $^{39}\textrm{Ar}$ spectrum in the ROI & $f_{2}$ & 0.20 $\pm$ 0.05 & Fixed & 1.1\\
Cut efficiency for double $^{39}\textrm{Ar}$ & $\epsilon_{2}$ & 0.9099 $\pm$ 0.0033 & Fixed & 1.8\\
\hline
Fraction of triple $^{39}$Ar spectrum in the ROI & $f_{3}$ & 0.19 $\pm$ 0.05 & Fixed & 0.01\\
Cut efficiency for triple $^{39}$Ar & $\epsilon_{3}$ & 0.860 $\pm$ 0.039 & Fixed & 0.19\\
\hline
Fraction of $^{39}$Ar-Cherenkov pile-up spectrum in the ROI & $f_{4}$ & 0.21 $\pm$ 0.05 & Fixed & 0.04\\
Cut efficiency for $^{39}$Ar-Cherenkov pile-up & $\epsilon_{4}$ & 1 & Fixed & --\\
\hline
Light yield corrections (constant) & & Run-dependent & N/A & 2.3\\
Light yield corrections (differential) & & Run-dependent & N/A & 5.1\\
Correlated triggers & & Run-dependent & N/A & 2.1\\
\hline
\end{tabular*}
\label{tab:Summary_table}
\end{table*}

\subsection{Results}
\label{sec:results}

\begin{figure} 
\includegraphics[width=\columnwidth]{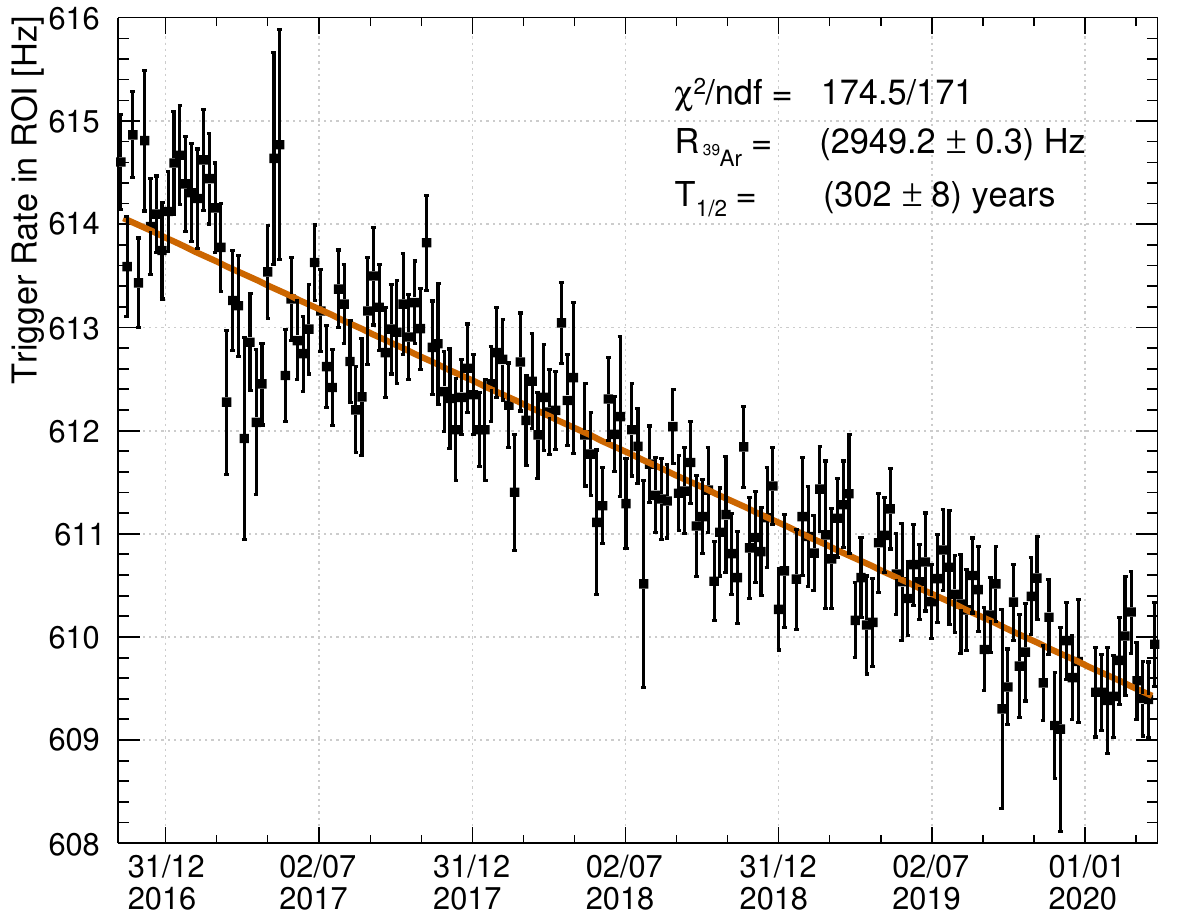}
\caption{The trigger rate fit for the full dataset. Each point here represents the rates averaged over a one week period. The best-fit values of R$_{\rm{39Ar}}$ and T$_{1/2}$ are shown with statistical uncertainties only.}
\label{fig:Rates}
\end{figure}

The trigger rate across the dataset is fit with the model in Eq.~\ref{eq:model} and shown in Figure~\ref{fig:Rates}. The measured trigger rate at the start of the dataset of $R_{\rm{39Ar}}~=~(2949.2 \pm 0.3_{\rm stat})$~{Hz}, comprised of mainly single $^{39}$Ar decays, but also double $^{39}$Ar pile-up, triple $^{39}$Ar pile-up, and $^{39}$Ar-Cherenkov pile-up, is consistent with the $^{39}$Ar decay rate measured in Ref. \cite{specificactivityAr39} given the systematic uncertainty on the $f_1$ parameter in this half-life measurement analysis. 

\begin{figure} 
    \includegraphics[width=\columnwidth]{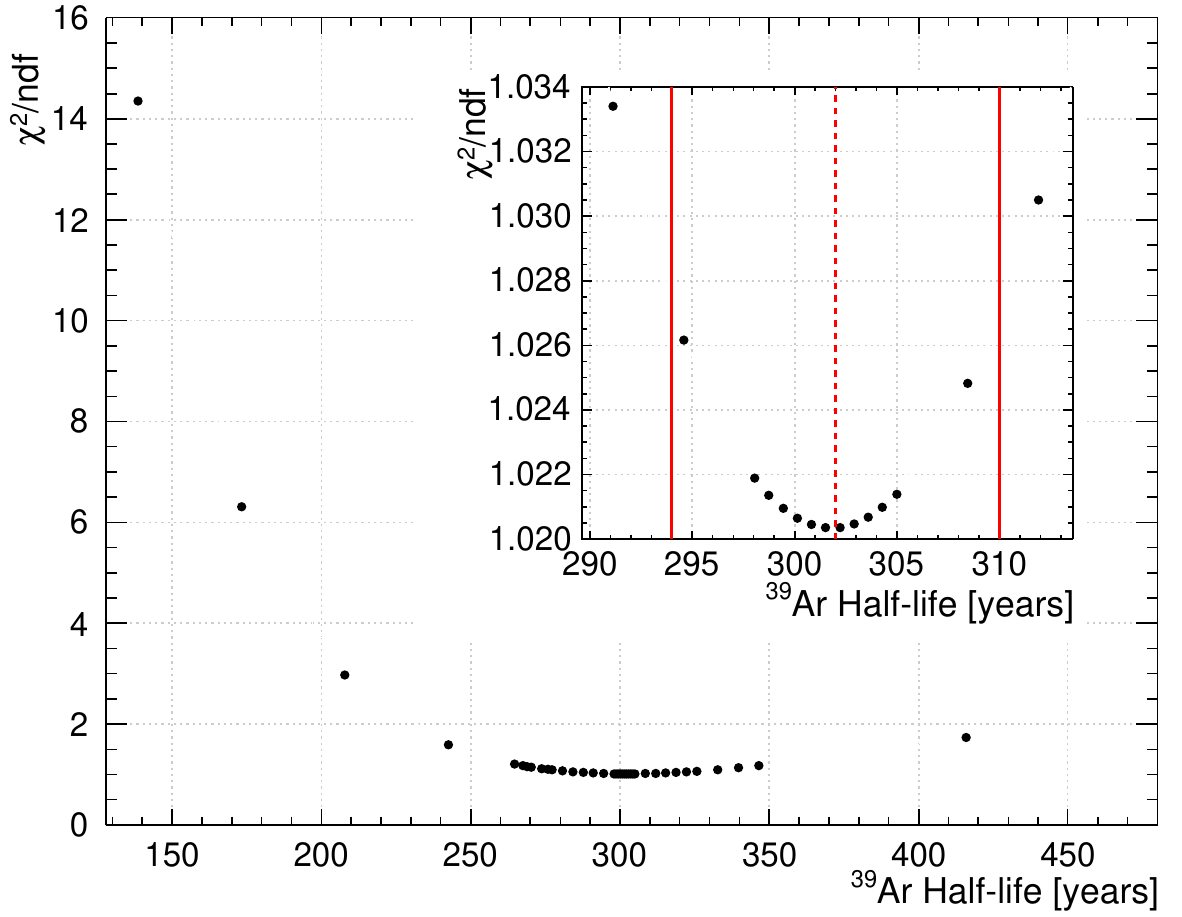}
    \caption{
    The $\chi^{2}$ to degrees of freedom (ndf = 171) ratio obtained by fixing T$_{1/2}$ in the fit to data. The insert plot shows the region around the minimum with the nominal T$_{1/2}$ (dashed line) $\pm 1\sigma$ (solid lines) region shown.}
    \label{fig:Rates_chisqndf}
\end{figure}

The fit is repeated for fixed values of T$_{1/2}$ between 140 and 415 years. Figure~\ref{fig:Rates_chisqndf} shows the reduced $\chi^2$ values from these fits.

\section{Systematic Uncertainties} 
\label{sec:systematic_uncertainties}
This section describes the individual contributions to the systematic uncertainty on the $^{39}$Ar half-life. These uncertainties are summarized in Table~\ref{tab:Summary_table}.
The total systematic uncertainty is calculated by adding uncertainties from each contribution in quadrature which results in a systematic uncertainty on T$_{1/2}$ of $\pm$~6 years. 

\subsection{Light Yield Corrections} 
\label{sec:uncertaity_w_time}

In order to assess the uncertainty on the half-life due to uncertainties in the light yield correction two components are considered: a constant offset component describes the effect of an absolute change in the light yield scale; a differential drift over time component describes the effect of continuous changes in the light yield scale across the dataset. To estimate these two components the uncertainty on the average light yield ratio, measured as the RMS of the distribution of uncertainties on $Y_j$ to be \mbox{$\Delta Y$ = 0.00076}, is considered in three ways. For the constant offset component a value of $\Delta Y$ is added to the average light yield ratio for each run in the dataset such that the correction described in Section~\ref{sec:lycorr} becomes $Y_j/Y_0 + \Delta Y$, and the fit to data is performed again. This results in a constant offset uncertainty of 2.3~years. The differential component is considered by correcting the average light yield ratio across the dataset as with the constant offset case where $\Delta Y$ is replaced with a linearly decreasing value in the range [$\Delta Y$,~0] and separately by replacing $\Delta Y$ with a linearly increasing value in the range [0,~$\Delta Y$]. For both cases the fit is re-done: this results in a differential drift over time uncertainty of 5.1~years.

As a cross-check on this method, the effect of each of the three calibration sources was considered relative to the average light yield. With each calibration source considered individually, the difference in the fraction of the $^{39}$Ar spectrum falling within the ROI relative to the spectrum fraction from the average light yield ratio is measured. This is done for each run in the dataset. This cross-check provides consistent results to the method above. The light yield corrections are the dominant source of systematic uncertainty on this half-life measurement.

\subsection{Energy Resolution} 
\label{sec:Energy_resolution_effect_on_tau}
Uncertainty due to the detector energy resolution is estimated by first fitting the $^{208}$Tl $\gamma$-ray peak for each run in the dataset and generating a distribution of the peak widths. The variation in this distribution is then taken as an additional input to the toy MC used to estimate the fraction $f_i$ of each event type's spectrum within the ROI (described in Section \ref{sec:FitModelValues}). The energy resolution is both increased and decreased by this variation, new fractions are measured for each case, and the fit to data is re-evaluated using the new fractions. Changing the detector resolution in this way has a negligibly small effect on the $^{39}$Ar half-life.

\subsection{ER Background Variation} 
\label{sec:ER_bg_variation}

As a cross-check method for the determination of $R_{\rm bg}$, contributions from the $^{232}$Th and $^{238}$U decay chains are examined. $^{208}$Tl rates, representative of the $^{232}$Th chain, are scaled based on the ratio of events in the peak to the low-energy tail events in the ROI. Similarly, $^{226}$Ra rates are examined to account for the $^{232}$U chain. The updated background rates are then added to the fit model and the fit performed again. The maximum observed difference in the $^{39}$Ar half-life from the nominal value is 0.15 years.

\subsection{Bin Width for Trigger Rate Averages}
To determine if there are any effects due to averaging the trigger rates over 7 day periods, the rates were separately averaged over 3, 5, 10, 14, 21, and 30 day periods and the fit performed again. The maximum observed difference corresponds to a change in the measured half-life of 1.0~years.

\subsection{Event Type Fractions in the ROI}

For each event type the fraction of events in the ROI are varied such that $f'_{i} = f_{i} \pm 0.05$. The uncertainty of $\pm$~0.05 is a conservative choice which considers uncertainty in the shape of the $^{39}$Ar spectrum and differences between theoretical models. One event type's fraction is varied at a time and the fit is performed again using each adjusted $f'_{i}$ value in Eq.~\ref{eq:model}. The differences between the nominal value and that from each new fit are taken as the systematic uncertainties and are listed in Table~\ref{tab:Summary_table}.

\subsection{Event Selection Efficiencies}

The run-by-run event selection efficiencies for each event type show small variations over the dataset. These variations are considered to estimate their effects on the $^{39}$Ar lifetime by adding them to the efficiencies in three ways and performing the fit again: (1)~the variation is kept constant; (2)~the variation increases from a value of 0 at the start of the dataset to the maximum value at the end; (3)~the variation decreases from the maximum value at the start of the dataset to 0 at the end. For each event type the largest change in the measured $^{39}$Ar half-life is observed for the third case. As a conservative approach these values are taken as the systematic uncertainties listed in Table~\ref{tab:Summary_table}.

\subsection{Correlated Triggers}
\label{sec:corrTrig}
A small fraction of events in DEAP-3600 are correlated with an earlier event on a time-scale that is greater than the data-cleaning cut. Light from these earlier events can leak into the trigger window of a subsequent event which affects the observed number of PE. The data-cleaning cut which rejects events occurring within $\delta t_{\rm cut}$ = 70~$\mu$s of an earlier event does not remove all correlated triggers, though a cut of greater than 70~$\mu$s would reduce the available statistics.

The effect of the correlated triggers is estimated on a run-by-run basis by fitting an exponential curve to the distribution of the time between triggers. The ratio of the data to the fit for each run's distribution is evaluated at a time of 70~$\mu$s. A linear fit to these values across the dataset is used to estimate the maximum expected effect on the change in trigger rate over time of 0.03 Hz. This corresponds to a 0.7\% change in the measured half-life, equal to 2.1~years.

\section{Conclusion}
\label{sec:conclusion}

A direct measurement of the half-life of $^{39}$Ar has been performed using data collected by the DEAP-3600 detector over a period of 3.4 years.  The statistical and systematic uncertainties are on the same order. The systematic uncertainty is dominated by effects from the light yield corrections. This measurement gives a value of the $^{39}$Ar half-life of 
\begin{equation*}
    T_{1/2} = (302 \pm 8_{\rm stat} \pm 6_{\rm sys})\ {\rm years}.
\end{equation*}
This new measurement is in tension with the half-life evaluated by NDS, with a $p$-value of 0.008 for the two measurements to have the same central value.

When considered with the specific activity measured in Ref.~\cite{specificactivityAr39} this half-life measurement corresponds to a decay constant of $\lambda$~=~\SI[separate-uncertainty]{7.4(0.4)e-11}~s$^{-1}$. The relative abundance of $^{39}$Ar in atmospheric argon is determined to be \SI[separate-uncertainty]{8.6(0.4)e-16}{}. In addition to impacting measurements sensitive to this isotope's half-life, such as studies of meteorites, this result is relevant for future experiments using atmospheric argon.

\begin{acknowledgements}

We thank the Natural Sciences and Engineering Research Council of Canada (NSERC),
the Canada Foundation for Innovation (CFI),
the Ontario Ministry of Research and Innovation (MRI), 
and Alberta Advanced Education and Technology (ASRIP),
the University of Alberta,
Carleton University, 
Queen's University,
the Canada First Research Excellence Fund through the Arthur B.~McDonald Canadian Astroparticle Physics Research Institute,
Consejo Nacional de Ciencia y Tecnolog\'ia Project No. CONACYT CB-2017-2018/A1-S-8960, 
DGAPA UNAM Grants No. PAPIIT IN108020 and IN105923, 
and Fundaci\'on Marcos Moshinsky,
the European Research Council Project (ERC StG 279980),
the UK Science and Technology Facilities Council (STFC) (ST/K002570/1 and ST/R002908/1),
the Leverhulme Trust (ECF-20130496),
the Russian Science Foundation (Grant No. 21-72-10065),
the Spanish Ministry of Science and Innovation (PID2022-138357NB-C2) and the Community of Madrid (2018-T2/TIC-10494), 
the International Research Agenda Programme AstroCeNT (MAB/2018/7)
funded by the Foundation for Polish Science (FNP) from the European Regional Development Fund,
and the Polish National Science Centre (2022/47/B/ST2/02015).
Studentship support from
the Rutherford Appleton Laboratory Particle Physics Division,
STFC and SEPNet PhD is acknowledged.
We thank SNOLAB and its staff for support through underground space, logistical, and technical services.
SNOLAB operations are supported by the CFI
and Province of Ontario MRI,
with underground access provided by Vale at the Creighton mine site.
We thank Vale for their continuing support, including the work of shipping the acrylic vessel underground.
We gratefully acknowledge the support of the Digital Research Alliance of Canada,
Calcul Qu\'ebec,
the Centre for Advanced Computing at Queen's University,
and the Computational Centre for Particle and Astrophysics (C2PAP) at the Leibniz Supercomputer Centre (LRZ)
for providing the computing resources required to undertake this work.

\end{acknowledgements}

\bibliographystyle{spphys}
\bibliography{bibliography}

\end{document}